\documentclass{IEEEcsmag}

\usepackage[colorlinks,urlcolor=blue,linkcolor=blue,citecolor=blue]{hyperref}

\usepackage{upmath}

\newcommand*{\nolink}[1]{%
  \begin{NoHyper}#1\end{NoHyper}%
}

\newcommand\notype[1]{\unskip}

\usepackage{listings}
\usepackage{xcolor}

\definecolor{codegreen}{rgb}{0,0.6,0}
\definecolor{codegray}{rgb}{0.5,0.5,0.5}
\definecolor{codepurple}{rgb}{0.58,0,0.82}
\definecolor{backcolour}{rgb}{0.95,0.95,0.92}

\lstdefinestyle{mystyle}{
    backgroundcolor=\color{backcolour},   
    commentstyle=\color{codegreen},
    keywordstyle=\color{magenta},
    numberstyle=\tiny\color{codegray},
    stringstyle=\color{codepurple},
    basicstyle=\ttfamily\footnotesize,
    breakatwhitespace=false,         
    breaklines=true,                 
    captionpos=b,                    
    keepspaces=true,                 
    numbers=left,                    
    numbersep=5pt,                  
    showspaces=false,                
    showstringspaces=false,
    showtabs=false,                  
    tabsize=2
}

\jvol{XX}
\jnum{XX}
\paper{8}
\jmonth{May/June}
\jname{IEEE Internet Computing}
\pubyear{2021}

\begin{document}

%\sptitle{Department: Head}
%\editor{Editor: Name, xxxx@email}

\title{Modeling Digital Twin Data and Architecture: A Building Guide with FIWARE as Enabling Technology}

\author{Javier Conde}
\affil{Departamento de Ingeniería de Sistemas Telemáticos, Escuela Técnica Superior de Ingenieros de Telecomunicación, Universidad Politécnica de Madrid}
\author{Andr\'es Munoz-Arcentales}
\affil{Departamento de Ingeniería de Sistemas Telemáticos, Escuela Técnica Superior de Ingenieros de Telecomunicación, Universidad Politécnica de Madrid}
\author{Álvaro Alonso}
\affil{Departamento de Ingeniería de Sistemas Telemáticos, Escuela Técnica Superior de Ingenieros de Telecomunicación, Universidad Politécnica de Madrid}
\author{Sonsoles López-Pernas}
\affil{Departamento de Ingeniería de Sistemas Telemáticos, Escuela Técnica Superior de Ingenieros de Telecomunicación, Universidad Politécnica de Madrid}
\author{Joaqu\'in Salvachúa}
\affil{Departamento de Ingeniería de Sistemas Telemáticos, Escuela Técnica Superior de Ingenieros de Telecomunicación, Universidad Politécnica de Madrid}

\markboth{Department Head}{Modeling Digital Twin Data and Architecture: A Building Guide with FIWARE as Enabling Technology}

\begin{abstract}
The use of Digital Twins in the industry has become a growing trend in recent years, allowing to improve the lifecycle of any process by taking advantage of the relationship between the physical and the virtual world. Existing literature formulates several challenges for building Digital Twins, as well as some proposals for overcoming them. However, in the vast majority of the cases, the architectures and technologies presented are strongly bounded to the domain where the Digital Twins are applied. This article proposes the FIWARE Ecosystem, combining its catalog of components and its Smart Data Models, as a solution for the development of any Digital Twin. We also provide a use case to showcase how to use FIWARE for building Digital Twins through a complete example of a Parking Digital Twin. We conclude that the FIWARE Ecosystem constitutes a real reference option for developing DTs in any domain.
\end{abstract}

\maketitle

\chapterinitial {Digital Twins} (DTs) have emerged with the aim of improving  process  productivity. The Industrial Internet Consortium (\url{https://www.iiconsortium.org/}) defines a DT as ``a formal digital representation of some asset, process or system that captures attributes and behaviors of that entity suitable for communication, storage, interpretation or processing within a certain context''  \cite{Digital_transformation_in_industry_white_paper}. In other words, DTs constitute the virtual counterpart of physical entities in the collaboration between the virtual and physical world \cite{A_review_of_the_roles}.

In order to make the use of DTs effective, it is necessary to overcome a series of challenges \cite{The_Digital_Twin_Realizing_the_Cyber-Physical}. First, DTs require to process and model an immense amount of data in batch and in real-time. For this purpose, distributed systems based on cloud computing are needed. Moreover, data acquisition from heterogeneous sources, each with its own format, and the combination of these data sources constitute a significant challenge, as does defining data models and ontologies that represent any entity regardless of its origin. Therefore, communication and connection protocols among all elements that compound a DT infrastructure must be standardized to ensure interoperability. Furthermore, it is necessary to deal with security in all the systems involved in the data exchange, both physical and virtual. DTs involve a large number of components and, in some cases, the information is very sensitive. Consequently, there must be authentication and authorization systems to protect the data, which must be encrypted in all communications. With the aim of addressing these challenges, new technologies and concepts have emerged, constituting the basis of DTs. Some of these are Big Data, Machine Learning, Internet of Things (IoT), and Linked Data.

Substantial research has been devoted to solving some of the challenges that arise from the requirements of DTs.
For instance,  \cite{Digital_Twin_and_Internet_of_Things_Current} compared the  most extended proposals regarding  data model definition. Asset Administration Shell (AAS), defined by Plattform Industrie 4.0, is a proposal developed to exchange information between real-world assets and digital tools \cite{The_Asset_Administration_Shell}. The Open Data Protocol (OData), developed by Microsoft and standardized in its fourth version by the Organization for the Advancement of Structured Information Standards (OASIS), was proposed for building and consuming RESTful APIs \cite{OData_version_4}. The Next Generation Service Interfaces-Linked Data (NGSI-LD), standarized by ETSI, was developed for managing context information with support to geo-spatial, temporal and historical data \cite{Context_Information_Management}. The Web of Things (WoT) group, at W3C, tries to remove the complexity related to the number of IoT standards and technologies by proposing the WoT Thing Description (DT) for modeling IoT resources, termed Things \cite{Web_of_Things_WoT}. Each of the alternatives presented allows to model data using one ore more formats among the most widely used in the industry (e.g., XML, AutomationML, RDF, JSON, and JSON-LD). 

Furthermore, a number of works have proposed architectures related to the flow of data in a DT. These architectures are, in most cases, strongly bounded to the field where the DT is applied. However, most of them have one thing in common: they are layered-architectures focused on acquiring, modeling, processing, and consuming data. An example of this four-layer architecture is found in \cite{A_four-layer}. However, in some approaches, the referenced architectures are segregated into more layers \cite{A_Six_Layer_Architecture}, by unfolding the layers related to acquiring and modeling data, resulting into six separate layers. In contrast, some other approaches merge two layers together into one single layer, e.g., the data processing and consuming layers \cite{A_Digital_Twin_Architecture_Based_on_the_Industrial_Internet_of_Things}. Other differences lie in how each layer is implemented. There are works that propose architectures in which the communication between the physical world and the virtual one is established through events and actions processed by a Cognitive System \cite{A_Simulation-Based_Architecture_for_Smart}. Others, such as \cite{Data-centric_Middleware}, are based on a central middleware oriented to the exchange of data in real-time by implementing Publish-Subscribe patterns, or distributed architectures based on microservices that implement different functions of the middleware \cite{a_microservice_base_middleware}. Other works, such as \cite{C2PS:A_Digital}, show architectures based on cloud computing. In addition, there are many proprietary cloud computing solutions (e.g., Microsoft Azure, IBM, GE Digital, and SAP) \cite{Digital_Twin_in_the_IoT_Context}.

Prior works propose FIWARE as a middleware for managing data in a DT \cite{Digital_Twin_Data_Modeling_with_AutomationML,Internet_of_things_ontology_for_digital_twin}. FIWARE (\url{https://www.fiware.org/}) is a framework of Open Source components, known as Generic Enablers (GEs), for facilitating the development and implementation of smart solutions. FIWARE provides all the necessary tools for building DTs. First, it allows for the definition and management of core context data in real-time. It is containerized, highly scalable, and easy to implement and replicate. Moreover, it is supported by the European Commission and the Context Broker GE belongs to the Connecting Europe Facility (CEF) Building Blocks, connecting the so-called ``Powered by FIWARE'' projects with other recognized European projects. However, the works that propose FIWARE as a middleware, do not delve into how to implement a complete solution using its GEs, but rather limit the role of FIWARE to the integration of the FIWARE Context Broker (for managing context data) and the IoT Agents (for connecting IoT devices to FIWARE infrastructure). The objective of the present article is to extend the definition of a DT reference architecture based on the FIWARE GEs. As a proof of concept, and also as a guide to develop a DT using the FIWARE GEs, we present a use case to demonstrate how architectures and data models based on FIWARE meet the requirements defined in the literature that need to be met in order to implement DTs.

The article is structured as follows. In the next section we present the data modeling based on FIWARE NGSI, followed by a reference architecture based on the FIWARE GEs. Thereafter, these proposals are applied in a use case consisting of a Parking DT. Lastly, the conclusions of the article are presented as well as an outlook of future work.

\section{MODELING DT DATA IN FIWARE}

As we have established, one of the cornerstones of designing and developing a DT is modeling data. Data originate from heterogeneous sources, use various protocols and include their own data attributes, attribute types and relationships. Not only is it necessary to standardize the communication among DT components, but standardizing the data format that flows through these components is also required to ensure interoperability. In this regard, FIWARE provides mechanisms for modeling and managing data. It introduces the Next Generation Service Interface (NGSI) and the FIWARE RESTful Application Programmer Interface (API), allowing systems to access the current state of context and perform updates using the NGSI API (\url{https://fiware.github.io/specifications/ngsiv2/stable/}).

The main elements of NGSI data are context entities: representations of physical or logical objects. Each of the context entities has a set of attributes and metadata of attributes to represent its real-life properties. Based on context entities, NGSI provides a data model for context information, an interface for exchanging data, and an interface for exchanging information on how to obtain such data. All these core elements are encoded in JSON format.

There is also a version of the NGSI interface that supports Linked Data: the NGSI Linked Data (NGSI-LD). NGSI-LD is an evolution of NGSI that supports formal relationships, semantics and property graphs using JSON-LD as the encoding method. The main goal of Linked Data is creating a machine-readable web of data in which each entity and relationship has a Uniform Resource Name (URN), allowing to access new sources of information by forming graphs of knowledge. NGSI API interactions are made using HTTP/HTTPs requests. Clients use HTTP methods for creating, updating, deleting and querying context information. 

The FIWARE community has already standardized many data models, from different domains, known as FIWARE Smart Data Models (\url{https://github.com/smart-data-models/}), with the objective of being portable for different solutions. Nowadays, there are six domains related to smart cities, smart environment, smart sensoring, smart agrifood, smart energy, smart water, and cross sector. Other domains, such as robotics, will be added in near future. Smart Data Models are compatible with both NGSIv2 and NGSI-LD. They are proposed and maintained by the community, and users are able to extend them for their specific use case.

\section{MODELING DT ARCHITECTURE IN FIWARE}

\begin{figure}[h!]
\centerline{\includegraphics[width=0.5\textwidth]{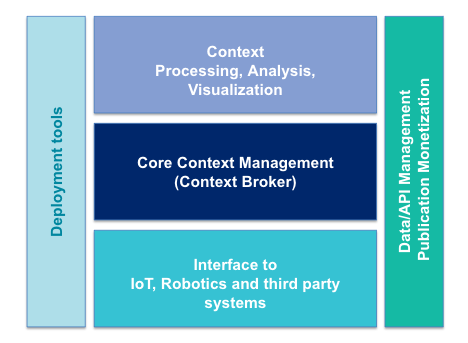}}
\caption{FIWARE GE catalogue. From: https://www.fiware.org/developers/catalogue/, with permission.}
\label{fig:catalogue}
\end{figure}

The centerpiece of any platform or solution ``Powered by FIWARE'' is the FIWARE Context Broker, bringing a cornerstone function in any smart solution: the need to manage, update and access context information. In the specific scope of DTs, the Context Broker provides a liaison between the physical and virtual world, acting as an intermediary in any exchange of information between the two domains or even within the same domain. Building around the FIWARE Context Broker, a rich suite of complementary components (the FIWARE Generic Enablers) are available, structured in three layers \nolink{(\textbf{Figure \ref{fig:catalogue}})}: (1) Interface with the Internet of Things, Robots and third-party systems, for capturing updates on context information in the physical world and translating required actuations; (2) Context Data/API management, enabling to perform updates, queries or subscribe to changes on context information; and (3) Processing, analysis, and visualization of context information, implementing the expected smart behavior of applications in the virtual world. The remaining of this section describes the proposed DT Reference Architecture, presenting all the FIWARE GEs involved and the relationship among them.

\subsection{Management of context data}

As we have stated before, in any ``Powered by FIWARE'' platform or solution, the Context Broker Generic Enabler is the core and mandatory component. The Context Broker GE enables to manage context information in a highly decentralized and large-scale manner. Currently, FIWARE offers four implementations of the Context Broker: Orion Context Broker, Orion-LD Context Broker, Scorpio Broker, and Stellio Context Broker. These GEs are capable of managing the different versions of the NGSI standard (\url{https://www.fiware.org/developers/catalogue/}). In this article, we focus our architecture proposal around the Orion Context Broker, the most extended GE, and the one endorsed by the European Commission as a CEF Building Block.

The \textbf{Orion Context Broker} (\url{https://fiware-orion.readthedocs.io/}) is the GE that manages context information through the implementation of a publish-subscribe pattern providing a NGSI interface. Clients can create context elements, query and update them, and subscribe to changes in context information for receiving notifications. Other elements interact with Orion through HTTP/HTTPs requests. Orion only saves the latest information state. Thus, if it is required to store the information history, other GEs have to be used.

\subsection{Data acquisition and persistence}
A DT needs to deal with multiple sources of data coming form physical devices. Hence, the data need to be transformed before they can be sent to the Context Broker. Fortunately, FIWARE  provides the IOT Agents GE to take care of this task. 

The \textbf{IoT Agents} (\url{https://github.com/FIWARE/catalogue/blob/master/iot-agents/README.md}) GE is a set of  software modules handling South IoT Specific protocols and North OMA NGSI interaction. These agents allow to work with the IoT devices that use communication protocols like LWM2M over CoaP, JSON or UltraLight over HTTP/MQTT, OPC-UA, Sigfox, or LoRaWAN. By using this component, the IoT devices will be represented in a FIWARE platform as NGSI entities in a Context Broker.

Additionally, the IoT agents GE allows to trigger commands to actuation devices just by updating specific command-related attributes in their NGSI entities representation at the Context Broker.

Furthermore, other sources of data can feed DTs. On the one hand, there are REST APIs, or Database systems that need to be queried periodically. On the other hand, there are systems like TCP or HTTP servers which send a continuous flow of data. These data, possibly coming from external applications, also need to be converted into NGSI and posted in the Context Broker. In order to deal with this type of systems that are not covered by the IoT Agents, FIWARE provides the Draco GE.  

\textbf{Draco} (\url{https://fiware-draco.readthedocs.io/}) is a dataflow management system based on Apache NiFi (\url{https://nifi.apache.org/}) that supports powerful and scalable directed graphs of data routing, transformation, and system mediation logic using a set of processors and controllers. The suite of processors and controllers provided in this GE allow to convert the incoming data into NGSI entities and attributes, required for their publication in the Context Broker.

Moreover, the Draco GE is used not only in the data acquisition block in the DT architecture, but also in the persistence part. Draco includes a set of processors in charge of persisting NGSI context data in third-party storage systems, allowing to create a historical view of the data posted in the CB.

\subsection{Data analysis}

In the scope of a DT, the data gathered from physical entities can be processed to extract relevant information or to draw conclusions that can influence, optimize and/or control any of the components of the DT without the need of human intervention. A DT equipped with such capabilities is known as an Intelligent DT  \cite{An_architecture_of_an_Intelligent_Digital_Twin_in_a_Cyber_Physical_Production_System}. In order to implement an Intelligent DT, Big data engines are required so as to handle large amounts of data.

The \textbf{Cosmos GE} (\url{https://fiware-cosmos-flink.readthedocs.io}) provides an interface for integrating two of the most well-known Big Data processing engines (namely, Apache Flink and Apache Spark) with the rest of the components of the FIWARE Ecosystem. This GE consists of a set of connectors that enable the reception of data from the Context Broker (through its subscription/notification feature) directly within data processing jobs running in these engines, allowing to process context data in real-time, as well as to send the result from this processing back to a given entity in the Context Broker so other components in the architecture can subscribe to changes in its attributes.

\subsection{Security architecture}

In the FIWARE Ecosystem, there are a number of GEs that manage authorization and authentication with regard to users and devices. In general, every request sent to a FIWARE GE has to be authenticated using a token. The \textbf{Keyrok GE} (\url{https://fiware-idm.readthedocs.io/}) is an Identity Management component based on OAuth 2.0. It is in charge of issuing the tokens for previously registered and authenticated subjects (users and devices). The \textbf{Wilma PEP Proxy GE} (\url{https://fiware-pep-proxy.readthedocs.io/}) acts as a Policy Enforcement Point (PEP), intercepting the aforementioned requests and validating the tokens to ensure the fulfillment of access control policies based on previously defined roles and permissions for subjects. Roles and permissions are defined and assigned by services' owners. In complex scenarios, where advanced policy definition is needed, the XACML-based Policy Decision Point (PDP) \textbf{AuthZForce GE} (\url{https://authzforce-ce-fiware.readthedocs.io/}) should be used. 

\section{USE CASE OF A DT MODELED WITH FIWARE}

In this section, we present a use case that depicts the guidelines to build a DT based on the FIWARE Ecosystem through an example of a Parking DT. This DT will be used to improve traffic in a parking lot. It will help customers find their vehicle through a 3D mobile application and predict the parking expected occupancy throughout a given day. The system collects real-time data from several sensors distributed around the parking space and data from an external API that provides weather information used in predictions.

\begin{figure*}
\centerline{\includegraphics[width=27pc]{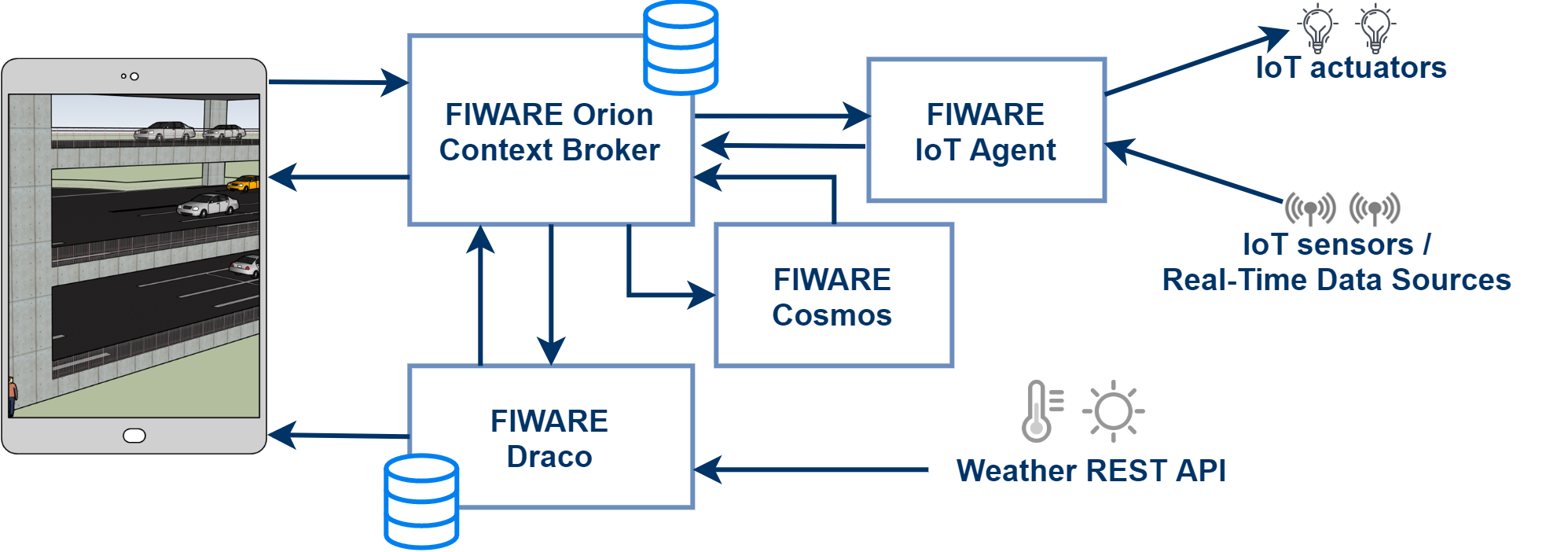}}
\caption{Simplified architecture of the Parking DT in which security GEs have been omitted for the sake of simplicity}
\label{fig:simplified_architecture}
\end{figure*}

The process of building the DT will be divided into four phases: data modeling, data acquisition, data consumption and security aspects. In \nolink{\textbf{Figure \ref{fig:simplified_architecture}}} the general architecture of the solution is presented.

\subsection{Data modeling}

For our system, we selected four different models, from the Smart Data Models catalog, to represent entities of type \texttt{OffStreetParking}, \texttt{ParkingSpot}, \texttt{Vehicle} and \texttt{WeatherForecast}. The first one models the whole parking; the second, a parking place; the third, a customer's vehicle; and the fourth, the weather forecast for the next hour. 

\subsection{Data acquisition}

Since the data originate from different sources, it is necessary to connect them through different protocols and adapt the information to the corresponding data models. We receive data from IoT sensors that report information of the vehicles and the place that they occupy in the parking. In this use case, sensors communicate through the Ultralight 2.0 protocol. An example of the payload when a vehicle parks is \texttt{id\textbar 123456\textbar t\textbar car\textbar p\textbar 51}. This payload represents a  \textbf{Vehicle} entity  of type \texttt{car} with plate number \texttt{123456} that is stationed in the parking spot number \texttt{51}. The respective IoT Agent receives this message, transforms it into the necessary entities in the NGSIv2 format and updates the context information sending three HTTP update requests to Orion. For the sake of simplicity, the samples presented in this use case do not include all the required attributes of the model (e.g. \texttt{location}). However, a complete solution must include them. The first request creates the entity of type \texttt{Vehicle} that represents this car:
\begin{lstlisting}
{
    "id": "vehicle:501",
    "type": "Vehicle",
    "vehicleType": "car",
    "vehiclePlateIdentifier": "123456"
}
\end{lstlisting}

The second one, updates the parking place 51, entity of model \texttt{ParkingSpot}, with a \texttt{status} occupied:

\begin{lstlisting}
{
    "id": "spot:51",
    "type": "ParkingSpot",
    "name": "51",
    "status": "occupied",
    "refVehicle": "vehicle:501",
    "refOffStreetParking": "parking:1"
}
\end{lstlisting}

Lastly, the third one decreases the \texttt{availableSpotNumber} of the \texttt{OffStreetParking} entity:

\begin{lstlisting}
{
    "id": "parking:1",
    "type": "OffStreetParking",
    "availableSpotNumber": 1449
}
\end{lstlisting}

The remaining data source is the weather API, which  provides  the weather forecast. Below is an example of these data:

\begin{lstlisting}
{
    "temp":27.50,
    "tempmin":27.08,
    "tempmax":27.60,
    "precipitation": 0.56,
    "wind":{
        "speed":1.5
    },
}
\end{lstlisting}

In this case, Draco retrieves the data periodically from the API and transforms them to fit the \texttt{WeatherForecast} model:

\begin{lstlisting}
{
    "id": "weatherForecast:2020-08-03T09",
    "type": "WeatherForecast",
    "validFrom": "2020-08-03T09:00:00.00Z",
    "validTo": "2020-08-03T10:00:00.00Z",
    "temperature": 27.50,
    "precipitationProbability": 0.56,
    "dayMaximum": {
        "temperature": 27.60
    },
    "dayMinimum": {
        "temperature": 27.08
    },
    "windSpeed": 1.5,
}
\end{lstlisting}

For this purpose, Draco uses two types of processors: one called InvokeHTTP, that makes the polling to the API, with a period of one minute, and another called JoltTransformJSON, which transforms the incoming JSON into the NGSIv2 format. Lastly, another InvokeHTTP processor sends an HTTP request to Orion and updates the context information. 

\subsection{Data consumption}

Orion provides the latest version of context information for the entities described and the other components subscribe to the changes in the entities they are interested in. Therefore, the 3D application will send an HTTP POST request to Orion in order to subscribe to changes in the status attribute of \texttt{ParkingSpot} entities. Consequently, when there is any change in the application the 3D frontend will receive HTTP requests from the Context Broker and it will be updated by adding or deleting the respective vehicles. Cosmos is subscribed to changes in the Context Broker too, specifically to the \texttt{WeatherForecast} entity and the \texttt{ParkingSpot} status attribute. With these data and some additional information (e.g., the hour of the day, the day of the week) Machine Learning algorithms can be applied to make useful real-time predictions for the parking such as the expected number of vehicles.

Finally, when changes in \texttt{ParkingSpot} entities take place, the IoT Agent receives a notification, processes it, and sends an action to update the light color of the smart bulb situated in the parking place. The chosen color code is red for closed parking place, yellow for occupied, and green for free. In this case, the bulb changes from green to yellow, and consequently, the payload of the Ultralight 2.0 sent message is \texttt{bulb:0051@light\textbar yellow}.

Storing historical data is of the utmost importance in our DT because historical data are used to train Machine Learning algorithms and periodically generate reports. Consequently, Draco subscribes to any changes in the context information and saves all data in a Mongo database using the ListenHTTP and NGSIToMongo processors.

\subsection{Security aspects}

\begin{figure}
\centerline{\includegraphics[width=18.5pc]{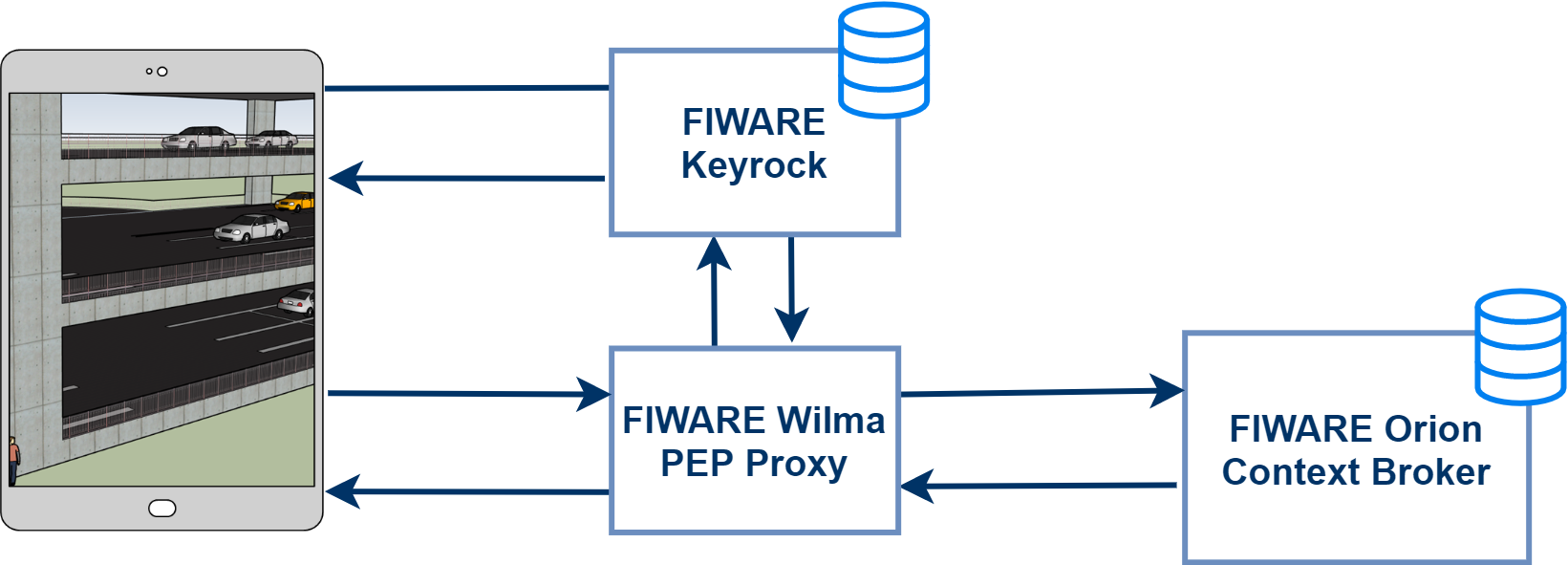}}
\caption{Extended architecture of the Parking DT with security GEs}
\label{fig:extended_architecture}
\end{figure}

Thus far, we have avoided talking about security in our Parking DT for the sake of simplicity, but it is indeed a topic that needs to be addressed. Confidentiality and integrity of data are fundamental aspects of DTs. Hence, it is necessary to secure the access to the Context Broker by managing roles and permissions. For that purpose, FIWARE provides two GEs that make this task easier. These are FIWARE KeyRock and FIWARE Wilma. We are going to focus on the access control of the 3D mobile application, but this proposal could be extended to the rest of the components in the architecture. \nolink{\textbf{Figure \ref{fig:extended_architecture}}} represents the extended architecture including security GEs.

We can distinguish three types of roles (i.e., admin, supervisor, and general user). Moreover, there are three main permissions (i.e., creating and deleting supervisors, updating the status of \texttt{ParkingSpot} entities, and retrieving the \texttt{ParkingSpot} entities). The admin has all three permissions, the supervisor can update the status of \texttt{ParkingPost} entities and retrieve them. The general user is only allowed to retrieve the \texttt{ParkingSpot} entities. These roles and permissions are created and assigned by the admin through the Keyrock interface or Keyrock REST API. Users that are not authenticated are redirected to Keyrock in order to sign in using any of the OAuth 2.0 grant flows available. The permissions are stated as an HTTP action an a resource. For example, the second permission is configured by the combination of action POST and resource \textit{/parkingSpot}.

FIWARE Wilma allows to secure the access to Orion, hiding its real address from external users. The parking users do not request information directly to the Context Broker, but instead send messages to Wilma, Wilma checks if the users have the required permissions and, if they do, it redirects the requests to Orion.

\section{CONCLUSIONS AND FUTURE WORK}

The success of DTs in today's industry, and consequently, the growth of their application in an increase number of fields, makes it necessary to standardize the way they are implemented, both from the components' side and that of the data flow. In this article we proposed FIWARE as a complete solution for building DTs, not only theoretically, but also practically through a use case of a Parking DT. On the basis of this work, it can be stated that FIWARE GEs and FIWARE Smart Data Models overcome the challenges formulated in the literature related to real-time and batch data consuming and processing, scalability, cloud computing, data modeling, and security. From our research we can conclude that FIWARE constitutes a legitimate reference option for the easy development of DTs of any kind. Moreover, being FIWARE GEs open source software, their evolution and adaptation to new requirements, technologies and standards is ensured thanks to FIWARE Community. 

As future work, the reference architecture should be validated with more use cases and the Smart Data Models repository needs to be expanded with additional domains. In the future we will propose the adaptation of this work to the NGSI-LD standard, using another Context Broker rather than Orion, and exploiting the advantages of Linked Data. Finally, as explained before, the proposed architecture is agnostic to the technology used in the DT visualization. However, it could be interesting to explore the benefits of using different approaches with regard to the visualization part.

\bibliographystyle{IEEEtran}
\bibliography{IEEEabrv, ref}

\begin{thebibliography}{10}
\providecommand{\url}[1]{#1}
\csname url@samestyle\endcsname
\providecommand{\newblock}{\relax}
\providecommand{\bibinfo}[2]{#2}
\providecommand{\BIBentrySTDinterwordspacing}{\spaceskip=0pt\relax}
\providecommand{\BIBentryALTinterwordstretchfactor}{4}
\providecommand{\BIBentryALTinterwordspacing}{\spaceskip=\fontdimen2\font plus
\BIBentryALTinterwordstretchfactor\fontdimen3\font minus
  \fontdimen4\font\relax}
\providecommand{\BIBforeignlanguage}[2]{{%
\expandafter\ifx\csname l@#1\endcsname\relax
\typeout{** WARNING: IEEEtran.bst: No hyphenation pattern has been}%
\typeout{** loaded for the language `#1'. Using the pattern for}%
\typeout{** the default language instead.}%
\else
\language=\csname l@#1\endcsname
\fi
#2}}
\providecommand{\BIBdecl}{\relax}
\BIBdecl

\bibitem{Digital_transformation_in_industry_white_paper}
\BIBentryALTinterwordspacing
M.~Buchheit, A.~Ferraro, C.~Lim, S.-W. Lin, J.~Morrish, and B.~Zarkout,
  ``{Digital Transformation in Industry White Paper},'' Industrial Internet
  Consortium, \notype, 2020. [Online]. Available:
  \url{https://www.iiconsortium.org/pdf/Digital_Transformation_in_Industry_Whitepaper_2020-07-23.pdf}
\BIBentrySTDinterwordspacing

\bibitem{A_review_of_the_roles}
E.~Negri, L.~Fumagalli, and M.~Macchi, ``{A review of the roles of Digital Twin
  in CPS-based production systems},'' \emph{{Procedia Manufacturing}}, vol.~11,
  pp. 939--948, 2017.

\bibitem{The_Digital_Twin_Realizing_the_Cyber-Physical}
T.~H.~J. Uhlemann, C.~Lehmann, and R.~Steinhilper, ``{The Digital Twin:
  Realizing the Cyber-Physical Production System for Industry 4.0},'' in
  \emph{{Proc. 24TH CIRP Conf. on Life Cycle Engineering}}, S.~Takata,
  Y.~Umeda, and S.~Kondoh, Eds., vol.~61, 2017, pp. 335--340.

\bibitem{Digital_Twin_and_Internet_of_Things_Current}
M.~Jacoby and T.~Usländer, ``{Digital Twin and Internet of Things—Current
  Standards Landscape},'' \emph{Applied Sciences}, vol.~10, no.~18, 2020.

\bibitem{The_Asset_Administration_Shell}
\BIBentryALTinterwordspacing
``{The Asset Administration Shell: Implementing digital twins for use in
  Industrie 4.0},'' {Plattform Industrie 4.0}, \notype, 2019. [Online].
  Available:
  \url{https://www.plattform-i40.de/PI40/Redaktion/EN/Downloads/Publikation/VWSiD
  V2.0.pdf}
\BIBentrySTDinterwordspacing

\bibitem{OData_version_4}
\BIBentryALTinterwordspacing
M.~Pizzo, R.~Handl, and M.~Zurmuehl, ``{OData Version 4.01. Part 1:
  Protocol},'' {Organization for the Ad-vancement of Structured Information
  Standards}, {OASIS Standard}, 2020. [Online]. Available:
  \url{https://docs.oasis-open.org/odata/odata/v4.01/odata-v4.01-part1-protocol.pdf}
\BIBentrySTDinterwordspacing

\bibitem{Context_Information_Management}
\BIBentryALTinterwordspacing
{European Telecommunications Standards Institute}, ``{Context Information
  Management (CIM); NGSI-LD API},'' {ETSI-GS-CIM-009}, 2020. [Online].
  Available: \url{https://www.etsi.org}
\BIBentrySTDinterwordspacing

\bibitem{Web_of_Things_WoT}
\BIBentryALTinterwordspacing
V.~Charpenay, T.~Kamiya, M.~McCool, S.~K{\"{a}}bisch, and M.~Kovatsch, ``{Web
  of Things (WoT) Thing Description},'' W3C, {W3C} Recommendation, Apr. 2020.
  [Online]. Available:
  \url{https://www.w3.org/TR/2020/REC-wot-thing-description-20200409/}
\BIBentrySTDinterwordspacing

\bibitem{A_four-layer}
S.~Malakuti, J.~Schmitt, M.~Platenius-Mohr, S.~Grüner, R.~Gitzel, and
  P.~Bihani, ``{A Four-Layer Architecture Pattern for Constructing and Managing
  Digital Twins},'' in \emph{Software Architecture, 13th European Conference,
  ECSA 2019}, ser. Lecture Notes in Computer Science, vol. 11681, 2019, pp.
  231--246.

\bibitem{A_Six_Layer_Architecture}
A.~Redelinghuys, K.~Kruger, and A.~Basson, ``{A Six-Layer Architecture for
  Digital Twins with Aggregation},'' in \emph{{Service oriented Holonic and
  Multi-Agent Manufacturing}}, T.~Borangiu, D.~Trentesaux, P.~Leitão,
  A.~Giret~Boggino, and V.~Botti~Navarro, Eds., 2020, pp. 171--182.

\bibitem{A_Digital_Twin_Architecture_Based_on_the_Industrial_Internet_of_Things}
V.~Souza, R.~Cruz, W.~Silva, S.~Lins, and V.~Lucena~Jr, ``{A Digital Twin
  Architecture Based on the Industrial Internet of Things Technologies},'' in
  \emph{Proc. IEEE Int.Conf. Consum. Electron. (ICCE)}, 2019, pp. 1--2.

\bibitem{A_Simulation-Based_Architecture_for_Smart}
T.~Gabor, L.~Belzner, M.~Kiermeier, M.~Beck, and A.~Neitz, ``{A
  Simulation-Based Architecture for Smart Cyber-Physical Systems},'' in
  \emph{{Proc. 2016 IEEE Inter. Conf. on Automotic Computing (ICAC)}}, 2016,
  pp. 374--379.

\bibitem{Data-centric_Middleware}
S.~Yun, J.-H. Park, and W.-T. Kim, ``{Data-centric Middleware based Digital
  Twin Platform for Dependable Cyber-Physical Systems},'' in \emph{{Proc. 2017
  Ninth Inter. Conf. on Ubiquitous and Future Networks (ICUFN 2017)}}, 2017,
  pp. 922--926.

\bibitem{a_microservice_base_middleware}
M.~Ciavotta, M.~Alge, S.~Menato, D.~Rover, and P.~Pedrazzoli, ``{A
  Microservice-based Middleware for the Digital Factory},'' \emph{Procedia
  Manufacturing}, vol.~11, pp. 931--938, 2017.

\bibitem{C2PS:A_Digital}
K.~M. Alam and A.~El~Saddik, ``{C2PS: A Digital Twin Architecture Reference
  Model for the Cloud-Based Cyber-Physical Systems},'' \emph{IEEE Access},
  vol.~5, pp. 2050--2062, 2017.

\bibitem{Digital_Twin_in_the_IoT_Context}
R.~Minerva, G.~M. Lee, and N.~Crespi, ``{Digital Twin in the IoT Context: A
  Survey on Technical Features, Scenarios, and Architectural Models},''
  \emph{Proceedings of the IEEE}, vol. 108, no.~10, pp. 1785--1824, 2020.

\bibitem{Digital_Twin_Data_Modeling_with_AutomationML}
G.~N. Schroeder, C.~Steinmetz, C.~E. Pereira, and D.~B. Espindola, ``{Digital
  Twin Data Modeling with AutomationML and a Communication Methodology for Data
  Exchange},'' \emph{IFAC-Papersonline}, vol.~49, no.~30, pp. 12--17, 2016.

\bibitem{Internet_of_things_ontology_for_digital_twin}
C.~Steinmetz, A.~Rettberg, F.~Ribeiro, G.~Schroeder, and C.~Pereira,
  ``{Internet of Things Ontology for Digital Twin in Cyber Physical Systems},''
  in \emph{{Proc. of SBESC 2018}}, 2018, pp. 154--159.

\bibitem{An_architecture_of_an_Intelligent_Digital_Twin_in_a_Cyber_Physical_Production_System}
B.~Ashtari~Talkhestani, T.~Jung, B.~Lindemann, N.~Sahlab, N.~Jazdi,
  W.~Schloegl, and M.~Weyrich, ``{An architecture of an Intelligent Digital
  Twin in a Cyber-Physical Production System},'' \emph{at -
  Automatisierungstechnik}, vol.~67, no.~9, pp. 762--782, 2019.

\end{thebibliography}

\begin {IEEEbiography} {Javier Conde}{\,} is currently pursuing a PhD in Telematics Engineering at UPM and a researcher in the Department of Telematics Engineering. His research interests lie in the fields of Open Linked Data, Digital Twins, Big Data and Machine Learning.
\end{IEEEbiography}

\begin{IEEEbiography}{Andr\'es Munoz-Arcentales} is currently working as a researcher in the Department of Telematics Engineering at UPM and studying a Ph.D. in Telematics Engineering at the same university, with a major research interests in Data Engineering, Data Usage Control and Machine Learning in Smart Spaces.
\end{IEEEbiography}

\begin{IEEEbiography}{\'Alvaro Alonso}{\,} is currently an Assistant Professor with the UPM. His research interests include Multi-videoconferencing Systems, Security Management in Smart Context scenarios and Public Open Data.
\end{IEEEbiography}

\begin{IEEEbiography}{Sonsoles L\'opez-Pernas}{\,} (GSM\textquotesingle19) is pursuing her PhD in Telematics Engineering at UPM. Since 2015, she has worked as a researcher in the Department of Telematics Engineering at UPM, participating in different projects such as FIWARE and eid4U.
\end{IEEEbiography}

\begin{IEEEbiography}{Joaqu\'in Salvach\'ua}{\,} is currently an Associated Professor with UPM. His research interests include Advanced Cloud and Edge architectures, Big Data infrastructure, Data Privacy and usage control, NoSql Databases, applications and identity in Blockchain.
\end{IEEEbiography}
\end{document}